\documentclass[useAMS,usenatbib,letterpaper]{mn2e}
\usepackage[totalwidth=480pt,totalheight=680pt]{geometry}
\usepackage{graphicx,amssymb}
\input psfig

\title[Non-Resonant Kepler Systems]
{Identifying Non-Resonant {\it Kepler} Planetary Systems}
\author[Veras \& Ford]{Dimitri Veras$^{1}$\thanks{E-mail:veras@ast.cam.ac.uk}, Eric B. Ford$^{2}$\\
$^{1}$Institute of Astronomy, University of Cambridge, Madingley Road, Cambridge CB3 0HA, UK\\
$^{2}$Astronomy Department, University of Florida, 211 Bryant Space Sciences Center, Gainesville, Florida 32111, USA}

\begin{document}

\date{Accepted 2011 November 01. Received 2011 October 26; in original form 2011 October 09}

\pagerange{\pageref{firstpage}--\pageref{lastpage}} \pubyear{2011} 

\maketitle

\label{firstpage}

\begin{abstract}
The {\it Kepler} mission has discovered a plethora of multiple transiting planet candidate 
exosystems, many of which feature putative pairs of planets near mean motion resonance 
commensurabilities.  
Identifying potentially resonant systems could help guide future observations
and enhance our understanding of planetary formation scenarios.
We develop and apply an algebraic method to determine which {\it Kepler} 2-planet
systems {\it cannot} be in a 1st-4th order resonance, given the current, 
publicly available data.  This method identifies when any potentially resonant 
angle of a system must circulate.
We identify and list 70 near-resonant systems which cannot
actually reside in resonance, assuming a widely-used formulation for 
deriving planetary masses from their observed radii and that these systems 
do not contain unseen bodies that affect the interactions of the observed planets.
This work strengthens the argument that a high fraction of exoplanetary systems 
may be near resonance but not actually in resonance.
\end{abstract}


\section{Introduction}

The {\it Kepler} mission has identified more planetary candidates
than the total number of extrasolar planets that were 
known before its launch \citep{boretal2011}.
This windfall of mission results enables investigators to perform
better statistical analyses to more accurately estimate properties of the exoplanet
population.  A dynamical question of particular interest is, ``what is the frequency
of resonant extrasolar systems?''  

The absence of resonance may be just as important, especially in systems with
tightly packed but stable configurations \citep{lisetal2011a}
and/or which might feature transit timing variations that span several orders
of magnitude \citep{foretal2011,veretal2011}.
Resonances may be a signature of a particular mode of planetary formation;
they are likely to be indicative of convergent migration in nascent
protoplanetary disks \citep{tholis2003,kleetal2004,papszu2005}.  Their 
absence might be indicative of
a dynamical history dominated by gravitational planet-planet scattering
\citep{rayetal2008}.  Alternatively, resonant statistics might best constrain
how disk evolution and gravitational scattering interface \citep{matetal2010,moearm2011}.

\cite{lisetal2011b} provide a comprehensive accounting of all planetary period 
ratios in {\it Kepler} systems, and present distributions and statistics linking
the periods to potentially resonant behavior.  Our focus here simply is to identify
which systems cannot be resonant.  A variety of mean motion resonances are known to 
exist in the Solar System and extrasolar systems.  However, for most of these cases, 
the masses, eccentricities and longitude of pericenters have been measured directly.  
Contrastingly, for {\it Kepler} systems, 
these parameters are constrained poorly, if at all.  Hence, analyzing resonant
{\it Kepler} systems poses a unique challenge.  

In order to address this situation, we present an analysis 
that i) treats the entire potential eccentricity phase space in most cases, 
ii) considers a relevant limiting
case of the unknown orbital angles, and iii) pinpoints the manner
in which planetary masses can affect the possibility of resonance.
The analysis is also entirely algebraic, allowing us to avoid numerical integrations and
hence investigate an ensemble of systems.  The results produce definitive claims 
about which systems {\it cannot} be in resonance, given the current data.  We 
consider the ensemble of {\it Kepler} two-planet systems which
may be near 1st-4th order eccentricity-based resonances, under the assumptions 
of coplanarity, non-crossing
orbits, and the nonexistence of additional, as-yet-unidentified planets
that would affect the interactions of the observed planets.  In Section 2, we 
explain our analytical model, and then apply it in 
Section 3.  We present our
list of non-resonant {\it Kepler} systems in Table 1, and briefly discuss 
and summarize our results in Section 4.

\section{Analytic Model}\label{criteria}

\subsection{Proving Circulation}\label{criteria}

The behavior of linear combinations of orbital angles from
each of the two planets determines if a system is in resonance or not.
``Libration'' is a term often used to denote oscillation of
a resonant angle, and ``circulation'' often refers to the absence of oscillation.
Resonant systems must have a librating angle; for mean motion 
resonances, the focus of this study, the librating angle incorporates
the positions of both planets.  This seemingly
simple criterion, however, belies complex behavior often seen
in real systems.

For example, \cite{braetal2004} 
illustrate how the resonant angle of a Neptunian Trojan can switch 
irregularly between libration and circulation, while \cite{fargol2006} 
illustrate how long periods of circulation can be punctuated by
short periods of libration.  
Characterizing libration versus circulation in extrasolar systems --
often with two massive extrasolar planets -- sometimes necessitates careful 
statistical measures \citep{verfor2010}.

The gravitational potential between two coplanar bodies orbiting a star can
be described by two ``disturbing functions,'' denoted by $R_w$, with $w=1,2$, which
are infinite linear sums of cosine terms with 
arguments of the form

\begin{equation}
\phi(t) = j_{1}  \lambda_1(t) + j_2 \lambda_2(t) + j_3 \varpi_1(t) + j_4 \varpi_2(t)
,
\label{phi}
\end{equation}

\noindent{where} $\lambda$ represents mean longitude, $\varpi$ represents
longitude of pericenter, 
the ``$j$'' values represent integer constants, and the subscript 
``$1$'' refers to the outer body while the subscript ``$2$'' refers to the 
inner body.  We choose this subscripting convention so
our equations conform with those in \cite{murder1999} and \cite{veras2007}.
The mean longitude is directly proportional to mean longitude at
epoch, denoted by $\epsilon_w$, such that 
$\lambda_w \equiv \varpi_w + M_w = \pi_w + \Omega_w + M_{w_0} + 
\int_{t_0}^{t} n_w(t') dt' = \epsilon_w + 
\int_{t_0}^{t} \mu_w^{(1/2)} a_w(t')^{(-3/2)} dt'$, where $M_w$ denotes
mean anomaly, $\pi_w$ denotes argument of pericenter,
and $n_w$ denotes mean motion.
The mean motion is related to its semimajor axis, $a_w$,  and mass through
Kepler's third law by $n_{w}^2 a_{w}^3 = \mu_w$, where 
$\mu_w = G \left( m_{0} + m_w \right)$, with $m_w$ representing the planet's
mass, $m_{0}$ the central body's mass, and $G$ the universal 
gravitational constant.  

The time derivative of Equation (\ref{phi}) yields:

\begin{eqnarray}
\dot{\phi}(t) &=& j_1 \mu_{1}^{\frac{1}{2}} a_{1}^{-\frac{3}{2}}(t) + j_1 \dot{\epsilon}_1(t) + j_2 \mu_{2}^{\frac{1}{2}} a_{2}^{-\frac{3}{2}}(t) + j_2 \dot{\epsilon}_2(t) 
\nonumber
\\
&+& j_3 \dot{\varpi}_1(t) + j_4 \dot{\varpi}_2(t)
.
\label{phidot}
\end{eqnarray}

Lagrange's Planetary Equations \citep[e.g.,][pp. 251-252]{murder1999} relevant to Eq. (\ref{phidot}) are:

\begin{eqnarray}
\frac{d \epsilon_{w}}{dt} &=& -a_{w}^{\frac{1}{2}} A_{w,1} \frac{\partial R_w}{\partial a_w} + a_{w}^{-\frac{1}{2}} 
A_{w,2} A_{w,3} \frac{\partial R_w}{\partial e_w} 
,
\\
\frac{d \varpi_{w}}{dt} &=& a_{w}^{-\frac{1}{2}} A_{w,2} \frac{\partial R_w}{\partial e_w}
,
\label{lagagmod2}
\end{eqnarray}

\noindent{where} 

\begin{eqnarray}
A_{w,1} &=& 2\mu_{w}^{-\frac{1}{2}},  \\
A_{w,2} &=& \mu_{w}^{-\frac{1}{2}} e_{w}^{-1} {\left( 1 - e_{w}^2 \right)}^{\frac{1}{2}},  \\
A_{w,3} &=& 1 - {\left( 1 - e_{w}^2 \right)}^{\frac{1}{2}}.
\end{eqnarray}

The form of $R_w$ used dictates how to proceed. \cite{veras2007} 
expresses \citeauthor{ellmur2000}'s
(\citeyear{ellmur2000}) disturbing function as

\begin{equation}
R_{w} = a_{1}^{-1} \sum_{y=1}^{\infty} 
                 \left[ \sum_{p=1}^{\infty} C_{w}^{(y,p)} X^{(y,p)} \right]
                 \cos{\phi^{(y)}}
,
\label{R1}
\end{equation}

\noindent{}where $C_{w}^{(i,p)}$ is a function of the masses and semimajor axes, and  
$X^{(y,p)}$ is a function of the eccentricities and inclinations.  Both
of these auxiliary variables contain the detailed functional forms needed
to model individual resonances.

When inserted into Eqs. (\ref{phidot})-(\ref{lagagmod2}),
the disturbing function will yield:

\begin{equation}
\dot{\phi}^{(u)} = \sum_{y}  \left(D^{(y,u)} \cos{\phi^{(y)}} \right) + E^{(u)} 
,
\label{phibarag2}
\end{equation}

\noindent{where} $\dot{\phi}^{(u)}$ is a potentially resonant angle, 
$D^{(y,u)}$ is an explicit algebraic function of the masses
and orbital elements, and 

\begin{equation}
E^{(u)} = \sum_{w = 1}^2 j_{w}^{(u)} \mu_{w}^{\frac{1}{2}} a_{w}^{-\frac{3}{2}}.
\label{Edef}
\end{equation}

Hence, a particular angle $\phi^{(u)}$ cannot librate if 

\begin{equation}
\sum |D^{(y,u)}| < |E^{(u)}|.  
\label{criterion}
\end{equation}

Equation (\ref{criterion}) represents the
criterion for an angle to circulate. Note that no integrations are required to 
evaluate the criterion.  For systems which cannot be in resonance,
the maximum value of $\sum |D^{(y,u)}|/|E^{(u)}| \equiv \beta_u$ provides
an estimate of the proximity to a potential resonance.

The dependence these variables have on planetary mass
is important to understand for the implications for {\it Kepler} systems:
$\mu_1$ or $\mu_2$ appears in each of the terms in $E^{(u)}$ and $D^{(y,u)}$,
and are insensitive to the planetary masses, as long as these masses are
negligible compared to the star's mass.  Further, each term in
$D^{(y,u)}$ is linearly proportional to either $m_1$ or $m_2$.
Therefore, if $m_1$ and $m_2$ are both scaled by the same factor
of their radii in a common mass-radius relationship, then the relative 
strength of the terms in $D^{(y,u)}$ will vary by a factor 
proportional to the planetary radii ratio. The signs of these
terms depend on the resonance being studied.  Hence, for a particular
resonance, one may determine which terms are additive, and 
obtain an explicit dependence on planetary mass.

\subsection{Hill Stability}

Unless two bodies are resonantly locked in a configuration that allows
for crossing orbits (such as Neptune and Pluto), the bodies will undergo
dynamical instability if their eccentricities are too great and their semimajor 
axis difference is too small.  Planets whose orbits never cross are said
to be Hill stable, and obey \citep{gladman1993}:

\begin{eqnarray}
&&\frac{1+ \eta_1 + \eta_2}{{\left( \eta_1 + \eta_1 \eta_2 + \eta_2 \right)^3}}
\left[
\eta_1 + \frac{\eta_2}{\alpha}
\right]
\nonumber
\\
&\times& \left[
\eta_2 \sqrt{\alpha \left( 1 - e_{2}^2 \right)}  + \eta_1 \sqrt{1 - e_{1}^2} 
\right]^2
>
\nonumber
\\
&&
1 + \frac{3^{\frac{4}{3}} \eta_1 \eta_2}{\left( \eta_1 + \eta_2 \right)^{\frac{4}{3}}}
  - \frac{\eta_1 \eta_2 \left( 11 \eta_2 + 7 \eta_1 \right)}{3 \left( \eta_1 + \eta_2 \right)^2}
,
\label{Hillst}
\end{eqnarray}

\begin{figure}
\centerline{
\psfig{figure=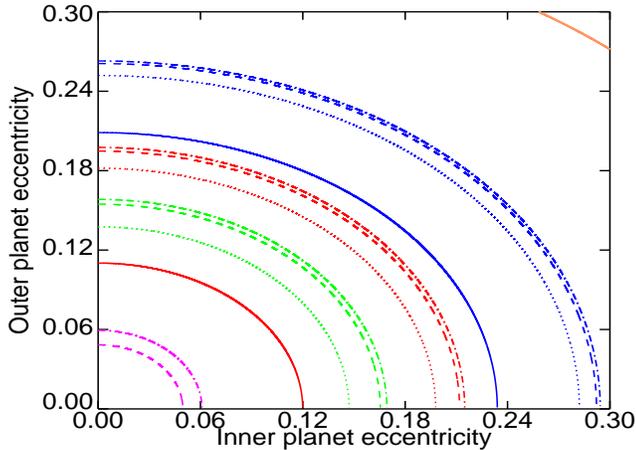,height=6cm,width=8.5cm} 
}
\caption{Hill stability eccentricity portrait.
Each curve (for the following planet/star mass ratios: 
solid lines - $10^{-3} m_{\odot}$; 
dotted lines - $10^{-4} m_{\odot}$;    
dashed lines - $10^{-5} m_{\odot}$;
dot-dashed lines - $10^{-6} m_{\odot}$)
bounds eccentricities at which two
planets may be Hill stable, for semimajor
axes ratios corresponding to the following
commensurabilities (in order of dashed lines
going outward from the origin):
$7$:$6$-magenta, $3$:$2$-green, $5$:$3$-red, $2$:$1$-blue, 
and $3$:$1$-salmon.  Stable systems lie below the curves,
and observed {\it Kepler} systems are assumed to be stable.
Note that the 
curves are not symmetrical about the origin.
The solid salmon curve indicates that nearly all
planets whose semimajor axis ratio is within a factor
of $\approx 2$ are subject to stability constraints
for $e_1, e_2 < 0.3$.
}
\label{smallevar}
\end{figure}

\noindent{where} $\eta_1 \equiv m_1/m_0$, $\eta_2 \equiv m_2/m_0$,
and $\alpha =  a_2/a_1$.
In principle, systems which do not satisfy Eq. (\ref{Hillst}) may be stable,
but generally this is true only for resonant systems.
Thus, we focus our efforts only on those systems which are provably stable.  
Fig. \ref{smallevar} displays level curves of Eq. (\ref{Hillst})
for different values of $\eta_1$ and $\eta_2$ 
($10^{-3} m_{\odot}$-solid, $10^{-4} m_{\odot}$-dotted, $10^{-5} m_{\odot}$-dashed, 
$10^{-6} m_{\odot}$-dashed-dot)
and at different commensurabilities of interest
($7$:$6$-magenta, $3$:$2$-green, $5$:$3$-red, $2$:$1$-blue, $3$:$1$-salmon), 
which each correspond to
the appropriate value of $\alpha$.  The range of eccentricities
plotted and considered in this study is well within the absolute
convergence limits of this disturbing function 
\citep{ferrazmello1994,sidnes1994} and corresponds roughly to a regime
where a fourth-order treatment (as in \citealt*{veras2007}) is accurate
to within $\approx 0.3^4 < 1\%$.

\section{Application to {\it Kepler} Systems}

\subsection{Method}

We use the publicly available {\it Kepler} 
database\footnote{ at http://archive.stsci.edu/kepler/planet\_candidates.html},
which provides directly measured values for planetary periods and 
estimated values of the stellar mass.
With these, we estimate the planetary semimajor axes
by assuming values for the planetary masses based on a mass-radius
relationship.  \cite{lisetal2011b}
estimate these values using:

\begin{equation}
m_1 = \left(\frac{R_1}{R_{\oplus}}\right)^b m_{\oplus}
\label{massb}
\end{equation}

\noindent{where} $b=2.06$.  A similar formula holds
for the other planet in the system.

We assume Eq. (\ref{massb}) to be true, and then 
determine the ``nearness'' of all two-planet 
systems to a mean motion commensurability through

\begin{equation}
\left( 1 - x/100\right) \left( j_1/j_2\right)^{2/3} 
< 
a_1/a_2  
< 
\left( 1 + x/100 \right) \left(j_1/j_2\right)^{2/3}
,
\label{prox}
\end{equation}

\begin{figure}
\centerline{
\psfig{figure=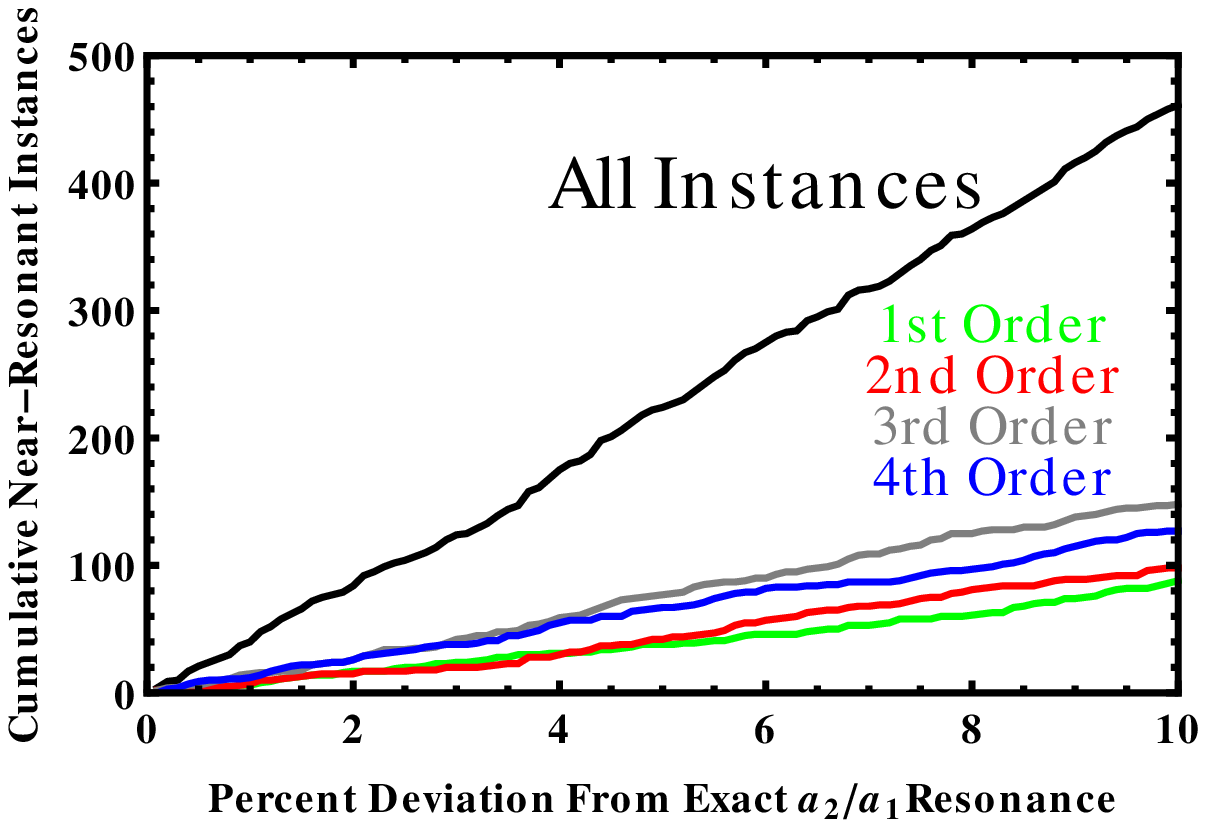,height=6cm,width=8cm} 
}
\centerline{
\psfig{figure=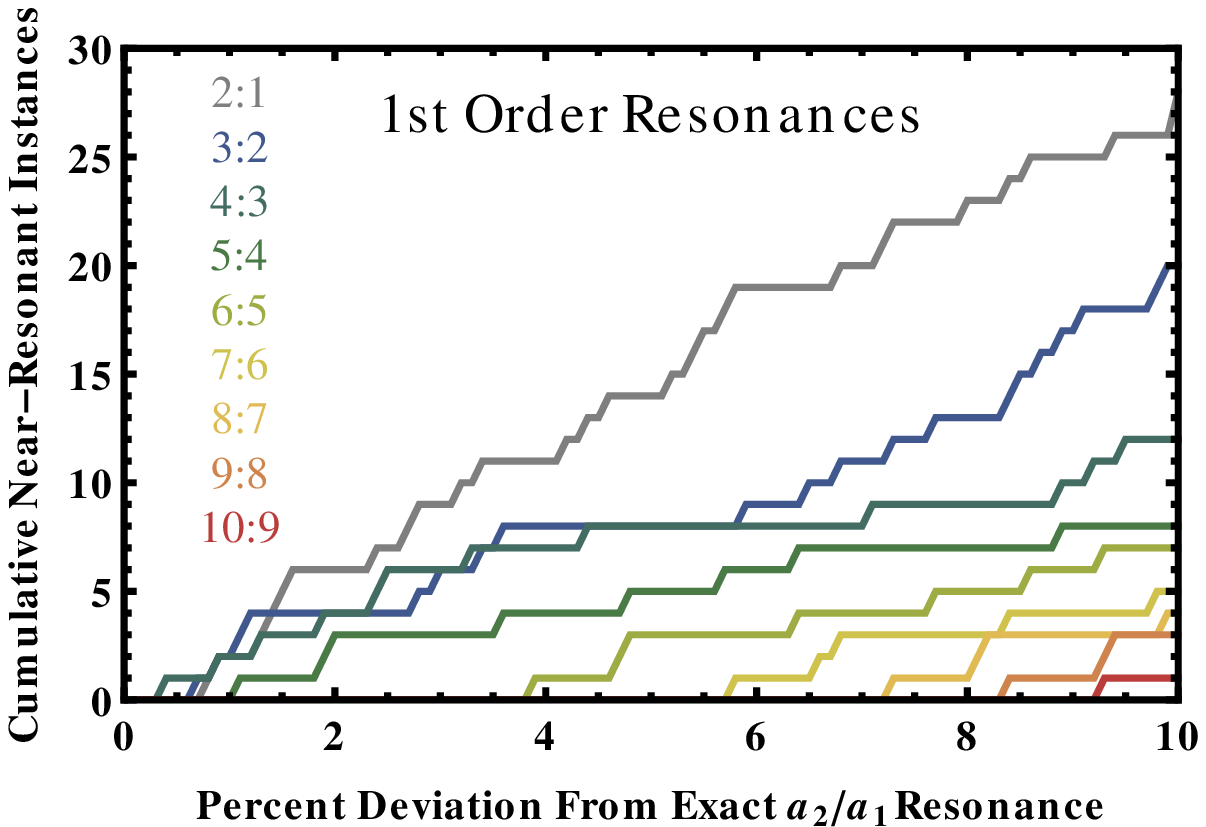,height=6cm,width=8cm} 
}
\centerline{
\psfig{figure=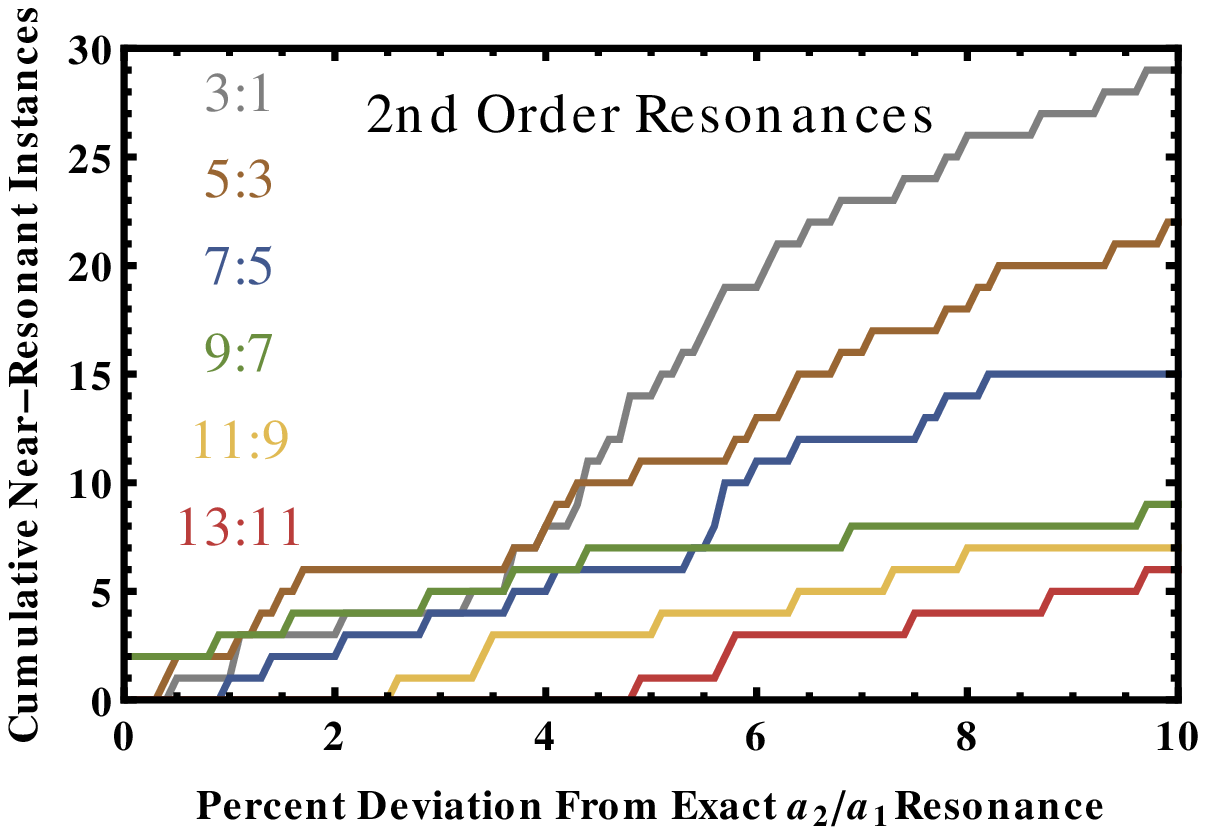,height=6cm,width=8cm} 
}
\caption{Number of near-resonant instances in 2-planet {\it Kepler} systems.
Plotted is the cumulative number of triplets $({\rm KOI},j_1,j_2)$ vs $x$, where
$x$ is defined in Eq. (\ref{prox}) and KOI refers to the
{\it Kepler} Object of Interest, or candidate exoplanet host star.
The upper panel plots all instances for 1st-4th order resonances,
and the other panels plot selected resonances of interest.
}
\label{freq}
\end{figure}

\noindent{where} $x$ effectively measures the percent offset from resonance
in semimajor axis space. Because we consider all potential resonances
up to fourth-order, the same planetary system may be close to several
resonances.  We refer to a ``near-resonant instance'' as a case
where Eq. (\ref{prox}) is satisfied.  We plot the cumulative frequency
of near-resonant instances as a function of $x$ in Fig. \ref{freq}.
Not included in the figure are $(j_1, j_2)$ multiplicities: higher-order 
harmonic angles which can be expressed as a linear combination of the angles 
with the lowest-order relatively prime values of $j_1$ and $j_2$, although 
these higher-order terms are included in the computation of $D^{(y,u)}$ 
\citep[see, e.g. Table 4 of][]{veras2007}.
The figure demonstrates the potentially wide variety of resonant
configurations which may be possible depending on one's
definition of ``near-resonant''.  We choose to be conservative and
include all ``near-resonant instances'' on the plots
in our analysis.

Having identified a potential resonance with
a given $j_1$ and $j_2$, and having
adopted $m_1$, $m_2$, $a_1$ and $a_2$,
we then sample the entire eccentricity phase
space for which the planets are Hill stable.
In the few cases where Hill stable systems
admit $e_1 > 0.3$ and/or $e_2 > 0.3$, we limit
the eccentricities to these values for 
accuracy and convergence considerations, 
as described earlier.  At each of 31 evenly-spaced
values of $e_2$ from $0$ to $0.3$, we compute
the maximum Hill stable value of $e_1$, and then 
sample 31 evenly-spaced values of $e_1$ from 0
to that Hill maximum value.  Finally, 
we compute $\beta_u = \sum_y |D^{(y,u)}|/|E^{(u)}|$ for
every possible disturbing function argument 
for $j_1,|j_2| > 0$ which satisfy the d'Alembert 
relations\footnote{
The secular arguments $j_1 = j_2 = 0$ up to 4th-order are
included in the computation of $D^{(y,u)}$ even though
we test only for circulation of angles which can lead
to mean motion resonance.}.
If no value of $\beta_u$ exceeds unity, then we flag 
the system as non-resonant and record the highest
value of $\beta_u$ as a proxy for closest proximity
to resonance.

\subsection{Results}

Of the 116 {\it Kepler} systems with two transiting candidates for which
public data is available, 94 have planets with periods
similar enough to each other to be considered
``near-resonant'' for at least one 1st-4th order
resonance with $x \le 10$.  These 94 systems
registered a total of 465 near-resonant instances
for $j_1 < 20$.  Of those 465, 313 cannot be resonant.
Only 4 systems (KOI \#s: 89, 523, 657, 738) have planets
which may be resonant for at least one instance and cannot
be resonant for at least one other, thereby restricting
the system's potential membership in some resonances.
For 20 systems (KOI \#s: 82, 111, 
115, 124, 222, 271, 314, 341, 543, 749, 787, 841, 870, 
877, 945, 1102, 1198, 1221, 1241, 1589), mean motion resonance
could not be ruled out for any of the near-resonant 
instances sampled.  Quantifying the likelihood of resonance
in these systems would require future detailed individual analysis.

We claim that the remaining 70 systems {\it cannot} be in resonance,
despite their close proximity to a commensurability.
We list these systems in Table 1, along with the number of their
near-resonant instances, their maximum value of $\beta$ and
the mean motion resonance this value corresponds to.
The larger the value of $\beta$, the closer the system
is to having parameters which could admit resonance.  In most
cases, $\beta < 0.5$, meaning that the systems
are well outside of resonance.  Notable exceptions 
are KOI 244, KOI 431, KOI 508, KOI 775, and KOI 1396,
three of which are close to the strong $2$:$1$ commensurability.
KOI 1151 is so close to so many commensurabilities because for that
system, $1/\alpha = 1.26$, meaning that the planets are tightly
packed and on the edge of stability.

\begin{table}
 \centering
  \caption{Non-Resonant 2-Planet {\it Kepler} Systems.}
  \begin{tabular}{@{}cccc@{}}
  \hline
    KOI & Number of & Closest & Maximum  \\
    Number & Near-Res Instances & $j_1$:$|j_2|$ & $\beta$ \\
 \hline
116 & 1 & $3$:$1$ & 0.078 \\
123 & 1 & $3$:$1$ & 0.018 \\
150 & 1 & $3$:$1$ & 0.027 \\
153 & 4 & $9$:$5$ & 0.28 \\
209 & 2 & $3$:$1$ & 0.22 \\
220 & 6 & $9$:$5$ & 0.039 \\
232 & 3 & $7$:$3$ & 0.016 \\
244 & 3 & $2$:$1$ & 0.67 \\
270 & 3 & $3$:$1$ & 0.0043 \\
279 & 4 & $9$:$5$ & 0.17 \\
282 & 1 & $3$:$1$ & 0.016 \\
284 & 2 & $3$:$1$ & 0.033 \\
291 & 1 & $4$:$1$ & 0.010 \\
313 & 3 & $7$:$3$ & 0.043 \\
339 & 1 & $3$:$1$ & 0.0043 \\
343 & 2 & $7$:$3$ & 0.12 \\
386 & 2 & $5$:$2$ & 0.16 \\
401 & 1 & $5$:$1$ & 0.031 \\
416 & 1 & $5$:$1$ & 0.024 \\
431 & 2 & $5$:$2$ & 0.60 \\
440 & 1 & $3$:$1$ & 0.040 \\
446 & 6 & $9$:$5$ & 0.18 \\
448 & 2 & $4$:$1$ & 0.038 \\
456 & 1 & $3$:$1$ & 0.048 \\
459 & 2 & $3$:$1$ & 0.042 \\
474 & 3 & $5$:$2$ & 0.020 \\
475 & 4 & $9$:$5$ & 0.23 \\
497 & 1 & $3$:$1$ & 0.31 \\
508 & 2 & $2$:$1$ & 0.84 \\
509 & 2 & $3$:$1$ & 0.042 \\
510 & 3 & $7$:$3$ & 0.035 \\
518 & 1 & $3$:$1$ & 0.028 \\
534 & 2 & $7$:$3$ & 0.37 \\
551 & 3 & $2$:$1$ & 0.25 \\
573 & 1 & $3$:$1$ & 0.11 \\
584 & 2 & $2$:$1$ & 0.083 \\
590 & 2 & $4$:$1$ & 0.0049 \\
612 & 3 & $7$:$3$ & 0.21 \\
638 & 2 & $3$:$1$ & 0.12 \\
645 & 2 & $3$:$1$ & 0.031 \\
658 & 6 & $9$:$5$ & 0.10 \\
672 & 3 & $5$:$2$ & 0.12 \\
676 & 1 & $3$:$1$ & 0.12 \\
691 & 5 & $9$:$5$ & 0.11 \\
693 & 5 & $9$:$5$ & 0.17 \\
700 & 1 & $3$:$1$ & 0.027 \\
708 & 3 & $7$:$3$ & 0.036 \\
736 & 2 & $3$:$1$ & 0.063 \\
752 & 1 & $5$:$1$ & 0.0049 \\
775 & 2 & $2$:$1$ & 0.78 \\
800 & 3 & $3$:$1$ & 0.023 \\
837 & 3 & $9$:$5$ & 0.035 \\
842 & 2 & $3$:$1$ & 0.090 \\
853 & 6 & $9$:$5$ & 0.19 \\
869 & 1 & $5$:$1$ & 0.051 \\
896 & 3 & $5$:$2$ & 0.14 \\
954 & 2 & $5$:$1$ & 0.0059 \\
1015 & 2 & $7$:$3$ & 0.085 \\
1060 & 2 & $5$:$2$ & 0.028 \\
1113 & 1 & $3$:$1$ & 0.030 \\
1151 & 15 & $9$:$7$ & 0.044 \\
1163 & 2 & $3$:$1$ & 0.015 \\
\hline
\end{tabular}
\end{table}

\begin{table}
 \centering
  \caption{...Table 1 continued}
  \begin{tabular}{@{}cccc@{}}
  \hline
    KOI & Number of & Closest & Maximum  \\
    Number & Near-Res Instances & $j_1$:$|j_2|$ & $\beta$ \\
 \hline
1203 & 3 & $7$:$3$ & 0.059 \\
1215 & 4 & $9$:$5$ & 0.084 \\
1278 & 1 & $4$:$1$ & 0.0081 \\
1301 & 1 & $3$:$1$ & 0.13 \\
1307 & 3 & $7$:$3$ & 0.049 \\
1360 & 2 & $5$:$2$ & 0.27 \\
1364 & 1 & $3$:$1$ & 0.19 \\
1396 & 6 & $9$:$5$ & 0.89 \\
\hline
\end{tabular}

\end{table}

We don't expect variations
in the masses of the {\it Kepler} planets to 
greatly affect the composition of Table 1, 
assuming that the value of $b$ from Eq. (\ref{massb})
does not vary by more than a few tenths
from 2.06.  The value of $\beta$ for a
particular system can hint at the potential
implications of mass variation.  In particular,
values of $\beta$ close to unity indicate
that the system is on the border of
potentially resonant behavior. For example,
for KOI 1396 ($\beta = 0.89$), if we
set $b = 1.7$, then the resulting planetary
masses could allow the system to harbor a (weak)
$9$:$5$ resonance.

\section{Conclusion and Implications}

We have identified 70 {\it Kepler} 2-planet near-commensurate systems 
which cannot be in an eccentricity-based mean motion resonance of
up to 4th-order.  These systems, may, in principle, achieve resonance
with crossing orbits or high ($e > 0.3$) eccentricities that could 
remain stable.  The criterion of 
Eq. (\ref{criterion}) is generally applicable to any 3-body
system suspected of harboring resonant behavior.
We caution that these results could be affected by the
presence of additional planets which have yet to be detected.

Systems which are provably non-resonant may be incorporated in
formation studies and detailed analyses of {\it Kepler} data. 
{\it Kepler} multi-planet systems may be preferentially 
clustered around particular resonances \citep{lisetal2011b}.
One possible explanation may be that convergent migration locks
planets in a mean motion resonance which later gets
broken by some additional perturbation.   We find that {\it Kepler} multi-planet
systems preferentially cluster around commensurabilities where
$|j_2|$ is low and rarely do so when $|j_2|$ is high.  
In particular, the number of {\it Kepler} systems near the $2$:$1$, $3$:$2$,
$3$:$1$ and $5$:$3$ commensurabilities is higher than what 
would be expected from a random distribution of {\it Kepler} planet
candidate semimajor axes.  Provably non-resonant
planets may also complement transit timing variation statistics,
as these variations take on distinctly different characteristics
for near-resonant and resonant systems \citep{veretal2011}.

The analysis in this work cannot be performed with {\it Kepler} systems 
that contain more than 2 planet candidates because i) more disturbing functions
must be incorporated, and hence the criteria for circulation becomes decidedly
more complex, and ii) analytical formulae for Hill stability no 
longer hold.  Figure 29 of \cite{chaetal2008} demonstrate that 
widely-separated pairs of planets in 3-planet systems whose orbits would 
be Hill stable in the 2-planet-only case may eventually become unstable.  
Even if a multi-planet system was assumed to be stable over a specified period 
of time and more disturbing functions were introduced, the resulting expansion 
of the phase space might render the computational cost of a similar 
algebraic analysis prohibitive compared to numerical integrations.
However, the investigation of the nonplanar 2-planet case 
with inclination-based resonances might be a fruitful avenue
to explore, especially because multiple transits detected by
{\it Kepler} may constrain the planets' mutual inclination, albeit weakly.

\section*{Acknowledgments}

We thank an anonymous referee, Geoff Marcy, and Darin Ragozzine 
for valuable comments.  This material is based on work supported by the
National Aeronautics and Space Administration under grant NNX08AR04G
issued through the {\it Kepler} Participating Scientist Program.


\label{lastpage}

\end{document}